\documentclass[preprint,aps,floatfix,showpacs]{revtex4}
\usepackage{dcolumn}
\usepackage{bm}

\usepackage[dvips]{graphicx,color}

\begin{document}


\title{Majority Orienting Model for the Oscillation of Market Price}

\author{TAKAHASHI, Hisanao}
\email{hisanao@ism.ac.jp}
\author{ITOH, Yoshiaki}%
\email{itoh@ism.ac.jp}\altaffiliation[Also at ]{
The Graduate University for Advanced Studies.
}
\affiliation{The Institute of Statistical Mathematics.\\
4-6-7 Minami-Azabu, Minato-ku, Tokyo 106-8569, Japan
}%

\begin{abstract}
The present paper introduces a majority orienting model in which the dealers' behavior changes based on the influence of the price to show the oscillation of stock price in the stock market.
We show the oscillation of the price for the model by applying the van der Pol equation which is a deterministic approximation of our model.
\end{abstract}

\pacs{89.65.Gh, 05.45.Tp, 02.50.Ey}
\maketitle

\section{Introduction}

The stock price is known to fluctuate in the real market and the trajectory of the stock price is considered to be a random walk, although the reason for the fluctuation remains an open question, as pointed out in the fundamental textbook of microeconomics \cite{Stiglitz}.
It is observed that the evolution of the price has some statistical properties or patterns that differs from a Gaussian random walk.
The present paper introduces a new model that is composed of three elements: mutation of dealers, majority rule and feedback by the price, as basic elements for the change of a stock price in a real market \cite{Cont Bouchaud,Iori,Lux Marchesi}.

The proposed model can be regarded as a simplified market in which the dealers' behavior (position) changes based on the influence of the price.
We show herein that in the model, the price oscillates under some condition by applying the van der Pol equation which is a deterministic approximation of the proposed model.

We introduce a ternary interaction model of a finite particle, which makes excursions that are similar to the Ising model, assuming a mutation to the other type for each particle \cite{Itoh Ueda, Hisanao em}.
Using the Ising model the analysis of the market has been discussed \cite{Aoki, Weidlich}, and modeling the financial market by some form of Ising structure of the interactions of agents has been achieved in several studies \cite{Cont Bouchaud,Iori}.
Based on the Ising model, spontaneous magnetization can only occur in the thermodynamic limit.
In a finite lattice, the system makes excursions from states with a uniformly negative magnetization through this intermediate mixed-phase state to states with a uniformly positive magnetization \cite{Binder}.

\section{Model}

Consider an urn model \cite{Karlin}.
Assume that there are two boxes, the plus box ($+$) and the minus box ($-$) and the number of particles contained in each box is $ N_+ $ and $ N_- $ respectively.
The total number of the particles is $ N = N_+ + N_- $, and every particle is numbered from 1 to $ N $.
The following step, composed of three substeps i), ii) and iii), is successively applied to particles in the two boxes.

\renewcommand{\theenumi}{\roman{enumi})}
\begin{enumerate}
\item {\it Mutation}:
One particle out of $ N $ particles is chosen at random and it moves from its box into the other box with probability $r$ and does not move with probability $1-r$, ($ 0\le r \le 1 $).

\item {\it Majority rule}:
We take three particles at random in each step.
If two of the particles taken are in the plus box and one is in the minus box, the one in the minus box is moved to the plus box and the price $ S $ is increased by 1, while, if two of the particles are in the minus box and one is in the plus box, the one in the plus box is moved to the minus box and the price $ S $ is decreased by 1.
If the three particles taken are in the plus box, no change occurs for the particles and the price $ S $ is increased by 3, while, if the three particles taken are in the minus box, no change occurs for the particles and the price $ S $ is decreased by 3.

\item {\it Feed back}:
We change the number of particles in the plus box $ N_+ $ (this may be equally applied to the minus box $ N_- $), with the probability which is proportional to the price $ S $.
That is, if $ S $ is a positive number, $ N_+ $ is decreased by 1 with probability $ S/N $, while, if $ S $ is negative, $ N_+ $ is increased by 1 with probability $ -S/N $.
The absolute value of $ S $ can be larger than $ N $ when $ r $ is large, but we only discuss in the case of $ |S| \le N $ in this paper.
This condition is almost valid when $ r \ge 0.65 $.
\end{enumerate}

Let each particle in the plus box represent a buy-position dealer, and each particle in the minus box represent a sell-position dealer.
Substep i) models the position of each dealer changing randomly in a real market and corresponds to mutation in population genetics.
This process may represent the random fluctuation (noise) in a real market.
Substep ii) models the change of the position of a dealer from sell to buy or from buy to sell after considering the behaviour of the other dealers \cite{Hisanao em}.
This process represents the majority orienting behaviour or herding behaviour of dealers in the real market.
This ternary interaction is one of the simplest interactions that we could think of for this type of urn model.
Substep iii) models the change of dealers' position under the influence of the price on the market.
If the price goes up, the buy-position dealers usually change their position to sell, whereas if the price goes down, the sell-position dealers may change their position to buy.

The ternary interaction in our model gives a linear term as we will show later (in Eq.~(\ref{d y/dt = x})), which gives the van der  Pol equation when combined with Eq.\ (\ref{emf d x/dt and d y/dt}).

\section{Van der Pol equation}

Let us represent $ N_+ $ and $ S $ at step $ s $ as $ N_+(s) $ and $ S(s) $ respectively.
Assuming that the duration of a step is $ \tau $, and the values of $ N_+(s) $, $ N_-(s) $ and $ S(s) $ are given, we have the following expected values:
\begin{eqnarray*}
	&&
	{\rm E}\Bigglb[ \frac{N_+(s+1)-N_+(s)}{\tau\, N}\Biggrb]
	\\
	&&=
	r\left\{-\frac{N_+(s)}{N}+\frac{N_-(s)}{N}\right\}
	\\
	&&
	\quad +3\frac{N_+(s)(N_+(s)-1)N_-(s)}{N(N-1)(N-2)}
	-3\frac{N_+(s) N_-(s)(N_-(s)-1)}{N(N-1)(N-2)}
	\\
	&&
	\quad -\lambda\frac{S(s)}{N}
\end{eqnarray*}
\begin{eqnarray*}
	&&
	{\rm E}\Bigglb[\frac{S(s+1)-S(s)}{\tau\, N}\Biggrb]
	\\
	&&=
	3\frac{N_+(s)(N_+(s)-1)(N_+(s)-2)}{N(N-1)(N-2)}
	+3\frac{N_+(s)(N_+(s)-1)N_-(s)}{N(N-1)(N-2)}
	\\
	&&\quad
	-3\frac{N_-(s)(N_-(s)-1)N_+(s)}{N(N-1)(N-2)}
	-3\frac{N_-(s)(N_-(s)-1)(N_-(s)-2)}{N(N-1)(N-2)},
\end{eqnarray*}
where $ \lambda $ is a parameter which we introduce for convenience in the analysis of the proposed model.

For sufficiently small $ \tau $ and large $ N $, approximating
\begin{eqnarray*}
	\frac{N_+(s)-N/2}{N} \mbox{~ by ~} x(t)
	\mbox{~~ and ~~}
	\frac{S(s)}{N} \mbox{~ by ~} y(t),
\end{eqnarray*}
we have the following system of ordinary differential equations as a deterministic approximation:
\begin{eqnarray}
	\label{emf d x/dt and d y/dt}
	\frac{d}{d t}x
	&=&
	- 2\,r\,x + 6\,x\left(\frac{1}{2}+x\right)\left(\frac{1}{2}-x\right) -\lambda \,y \\
	\frac{d}{d t}y
	&=&
	3\left(\frac{1}{2}+x\right)^3+3\left(\frac{1}{2}+x\right)^2\left(\frac{1}{2}-x\right)
	\nonumber
	\\
	& & 
	-3\left(\frac{1}{2}+x\right)\left(\frac{1}{2}-x\right)^2
	-3\left(\frac{1}{2}-x\right)^3
	\nonumber
	\\
	&=&
	6\, x,
	\label{d y/dt = x}
\end{eqnarray}
where $ -1/2 \le x \le 1/2 $.

\begin{figure}
(a)\hspace{7cm}\vspace{-.5cm}
	\begin{center}
	\rotatebox{90}{\hspace{1.cm}$ y(t) $}
	\includegraphics[height=3cm]{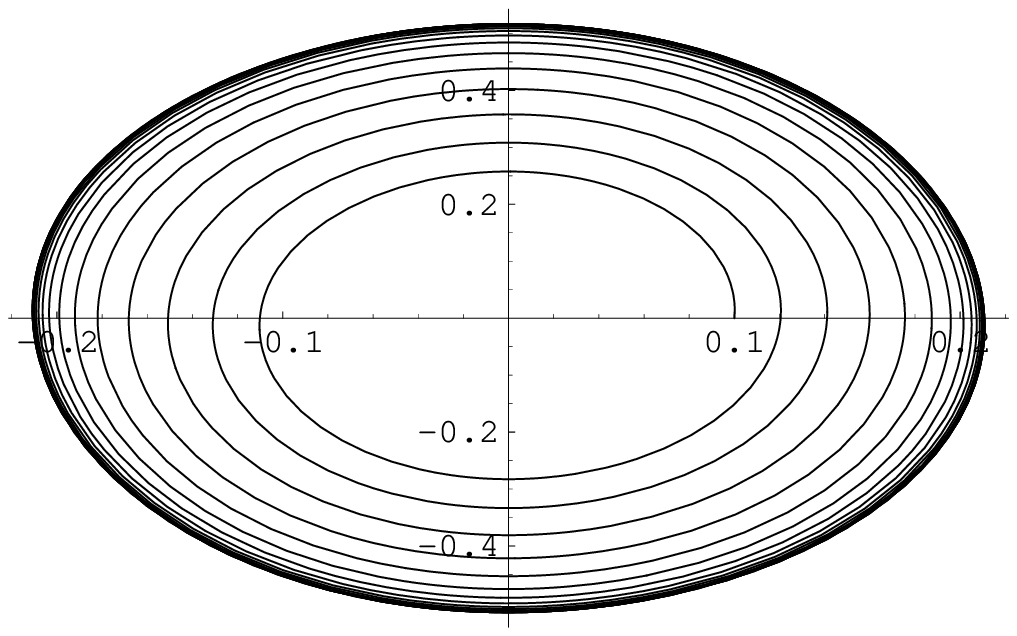}\\
	\qquad $ x(t) $\\
	\end{center}
(b)\hspace{7cm}\vspace{-.5cm}
	\begin{center}
	\rotatebox{90}{\hspace{1.cm}$ y(t) $}
	\includegraphics[height=3cm]{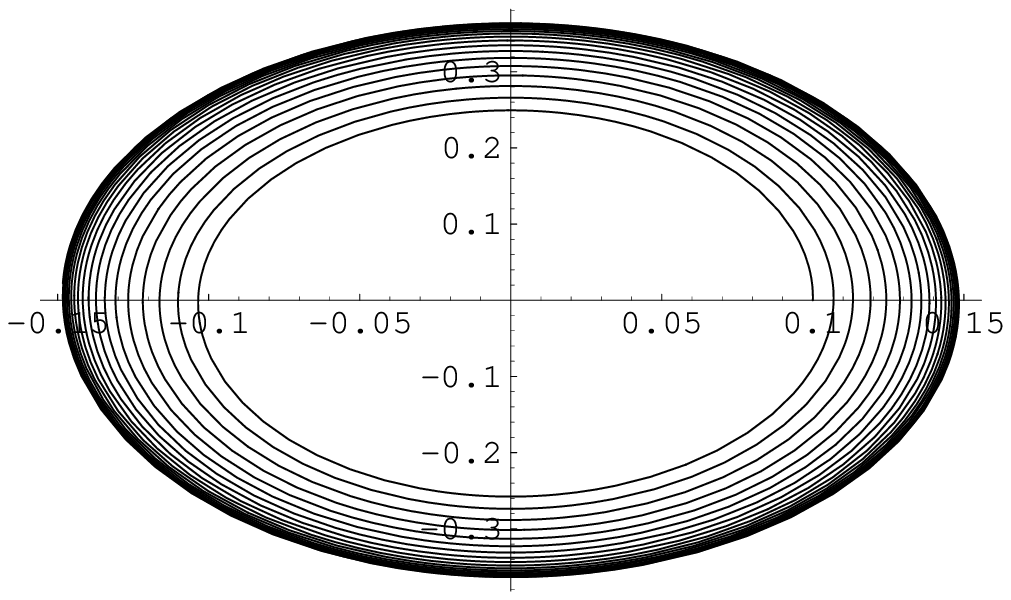}\\
	\qquad $ x(t) $\\
	\end{center}
(c)\hspace{7cm}\vspace{-.5cm}
	\begin{center}
	\rotatebox{90}{\hspace{1.cm}$ y(t) $}
	\includegraphics[height=3cm]{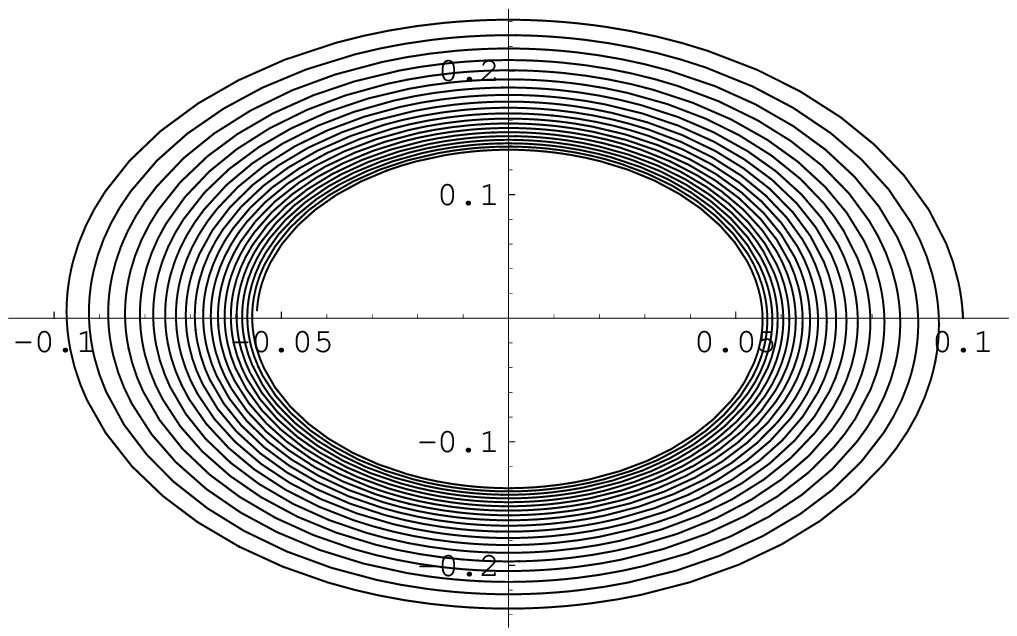}\\
	\qquad $ x(t) $\\
	\end{center}
(d)\hspace{7cm}\vspace{-.5cm}
	\begin{center}
	\rotatebox{90}{\hspace{1.cm}$ y(t) $}
	\includegraphics[height=3cm]{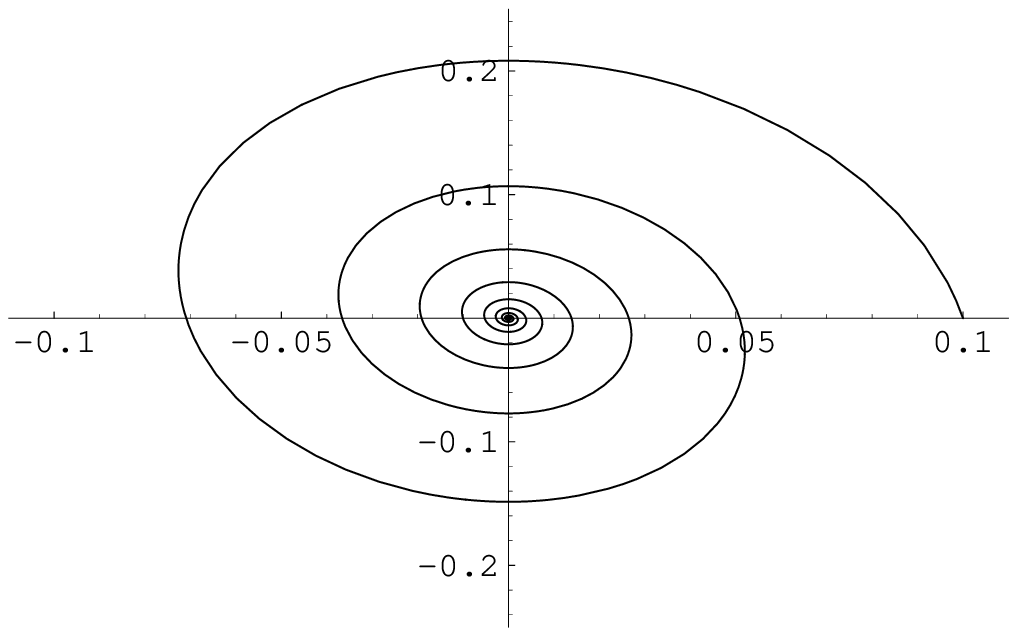}\\
	\qquad $ x(t) $\\
	\end{center}
	\caption[]
	{Trajectories of the price $ y(t) $ and the position $ x(t) $. (a) $ r = 0.65 $, (b) $ r = .70 $, (c) $ r = .75 $, and (d) $ r= 1.0 $. With the initial condition that $ x(0) = .1 $ and $ y(0) = 0 $.
	}
	\label{fig. trajectories of price and mind. deterministic}
\end{figure}

In Fig.\ \ref{fig. trajectories of price and mind. deterministic}, we show the trajectories of the above equations under the initial conditions
\[
	x(0)= 0.1\mbox{~~and~~}y(0)=0.
\]
The time interval of this simulation is 0.017.
From these figures, we find that when the mutation rate $ r $ is small, the price and the position are oscillating, while, when the mutation rate $ r $ is large, these damp to zero.
When $ r $ is too small, the trajectory goes to $ |x| > 1/2 $.

The following van der Pol equation \cite{Jackson,Ozaki} is obtained from Eqs. (\ref{emf d x/dt and d y/dt}) and (\ref{d y/dt = x}):
\begin{equation}
	\label{emf van der Pol}
	\frac{d^2}{dt^2}x
	-6\left(\frac{3-4\,r}{12}-3\,x^2\right)\frac{d}{dt}x
	+6\, \lambda \,x=0,
\end{equation}
when $ r < 3/4\ (\simeq 0.75) $.
The limit cycle was shown in Fig.~\ref{fig. trajectories of price and mind. deterministic} of the simulation.

\section{Simulation}

\begin{figure}
(a)\hspace{7cm}\vspace{-.5cm}
	\begin{center}
	\rotatebox{90}{\hspace{.4em}$ N_+(s)-N/2 $, $ S $}
	\includegraphics[height=3cm]{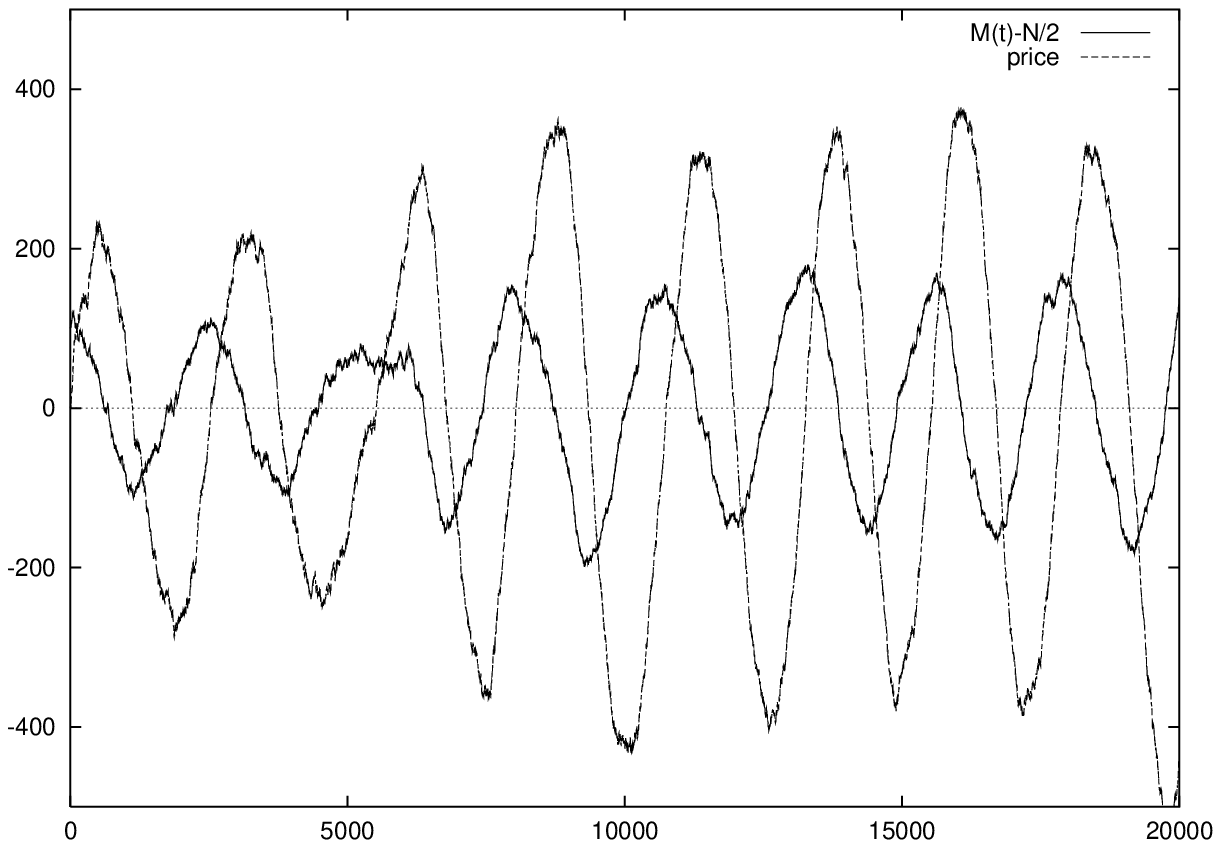}\\
	\qquad~ $ s $
	\end{center}
(b)\hspace{7cm}\vspace{-.5cm}
	\begin{center}
	\rotatebox{90}{\hspace{.4em}$ N_+(s)-N/2 $, $ S $}
	\includegraphics[height=3cm]{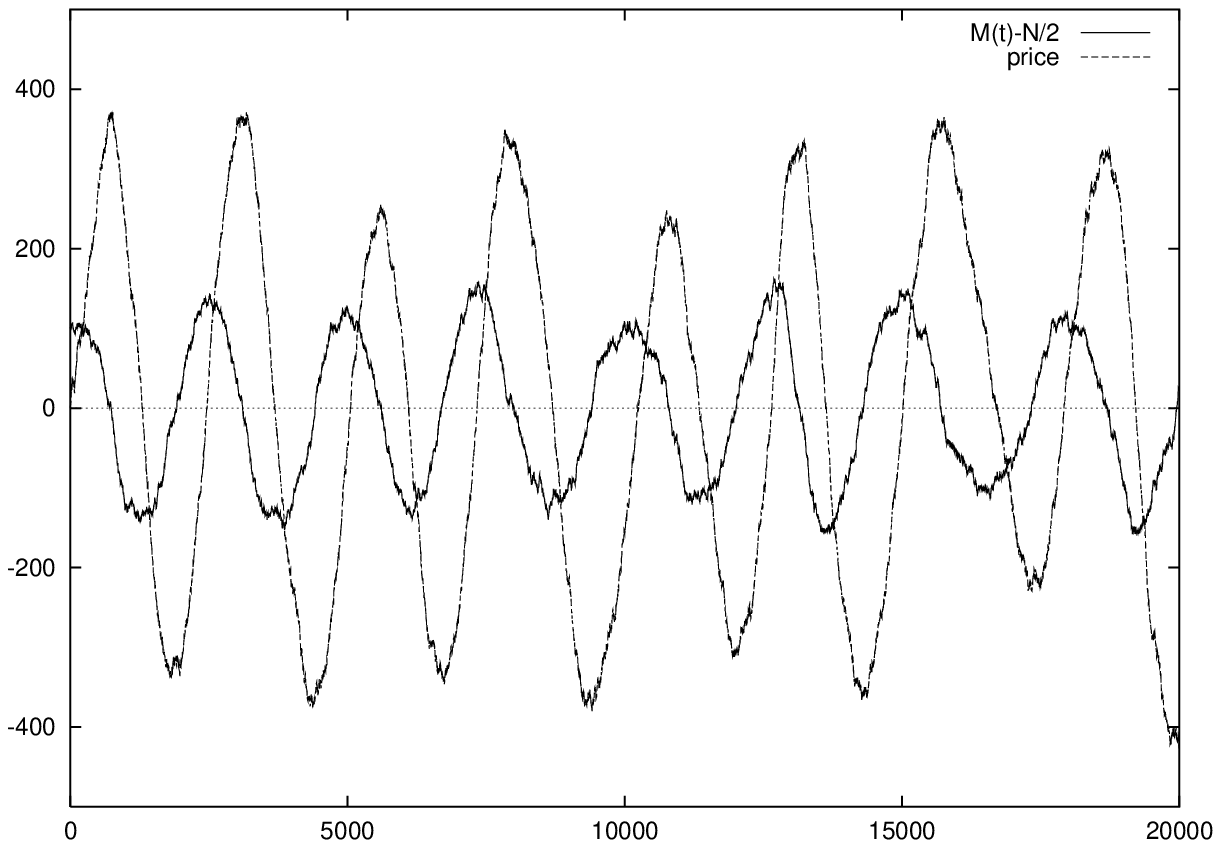}\\
	\qquad~ $ s $
	\end{center}
(c)\hspace{7cm}\vspace{-.5cm}
	\begin{center}
	\rotatebox{90}{\hspace{.4em}$ N_+(s)-N/2 $, $ S $}
	\includegraphics[height=3cm]{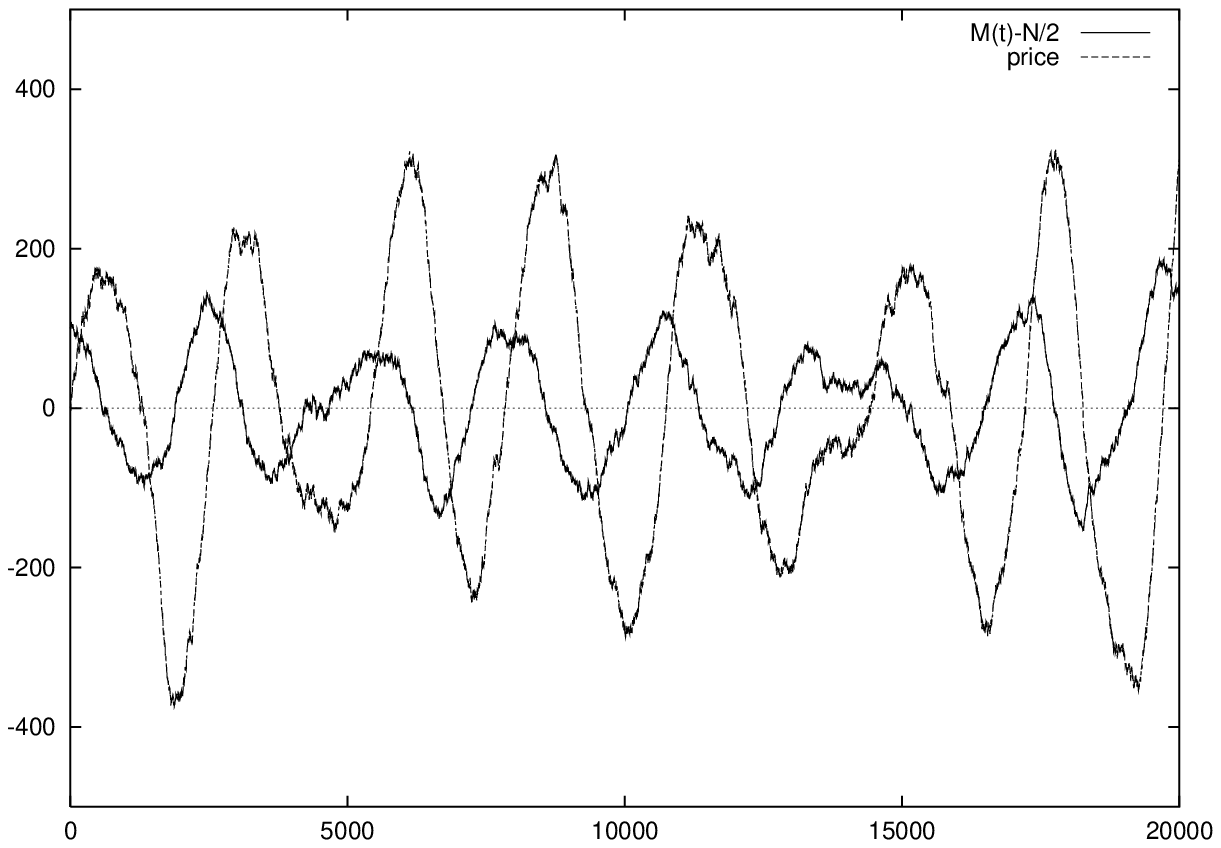}\\
	\qquad~ $ s $
	\end{center}
(d)\hspace{7cm}\vspace{-.5cm}
	\begin{center}
	\rotatebox{90}{\hspace{.4em}$ N_+(s)-N/2 $, $ S $}
	\includegraphics[height=3cm]{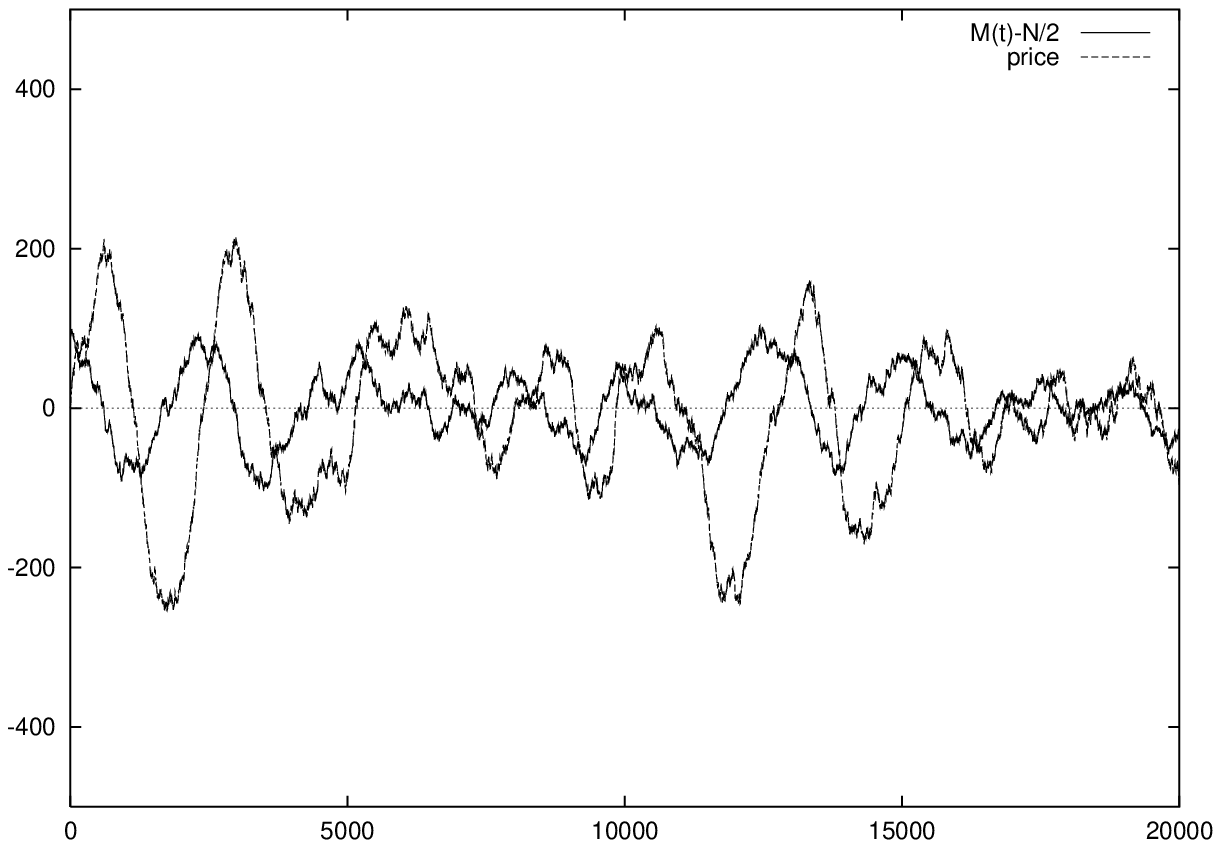}\\
	\qquad~ $ s $
	\end{center}
	\caption[]
	{Trajectories of the number of particles in plus box $ N_+(s) $ and the price $ S(s) $ for $ N = 1000 $. (a) $ r = .65 $, (b) $ r = .70 $, (c) $ r = .75 $, and (d) $ r= 1.0 $. 
	}
	\label{fig. trajectories of price and mind}
\end{figure}

In Fig.~\ref{fig. trajectories of price and mind}, we show the trajectories with four different parameters $ r $:
(a) $ r = .65 $, (b) $ r = .70 $, (c) $ r = .75 $, and (d) $ r= 1.0 $.
The number of particles $ N = 1000 $.
The number of particles in the plus box $ N_+(0) = 600 $, and the price number $ S(0) = 0 $.
The Mersenne Twister \cite{Mersenne} was used to generate random numbers.

We find from Fig.~\ref{fig. trajectories of price and mind} that when $ r < 3/4\ (=0.75) $, there is a limit cycle, as is known for the van der Pol equation.
When $ r > 3/4 $, the price tends to zero, which is realized from Eqs.~(\ref{emf d x/dt and d y/dt}), (\ref{d y/dt = x}) and (\ref{emf van der Pol}).
Considering the physical interpretation in this case, Eq.~(\ref{emf van der Pol}) has no energy supplying term but diffusion or friction terms, so that the system loses its energy, and the position of traders and the price tend to zero.
These results are consistent with the results from Fig.\ \ref{fig. trajectories of price and mind. deterministic}.

Looking carefully at Fig.~\ref{fig. trajectories of price and mind} (d), we observe a periodic pattern with oscillation.
This is because $ N_+ $ drifts away from $ N/2 $ in proportion to $ s $ due to the Brownian motion caused by the random fluctuations, and gradually $ S $ will become bigger, then $ N_+ -N/2 $ goes to zero following from Eq.~(\ref{emf van der Pol}), and these processes continue.

\begin{figure}
(a)\hspace{7cm}\vspace{-.5cm}
	\begin{center}
	\rotatebox{90}{\hspace{1.0cm}frequency}
	\includegraphics[height=3cm]{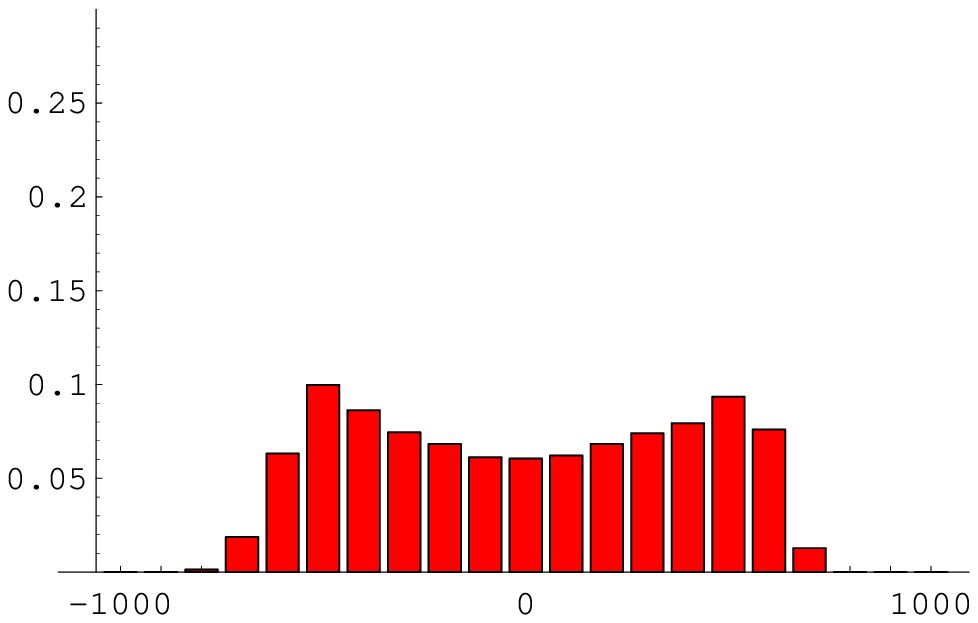}\\
	\vspace{-2mm}\qquad $ S $
	\end{center}
(b)\hspace{7cm}\vspace{-.5cm}
	\begin{center}
	\rotatebox{90}{\hspace{1.0cm}frequency}
	\includegraphics[height=3cm]{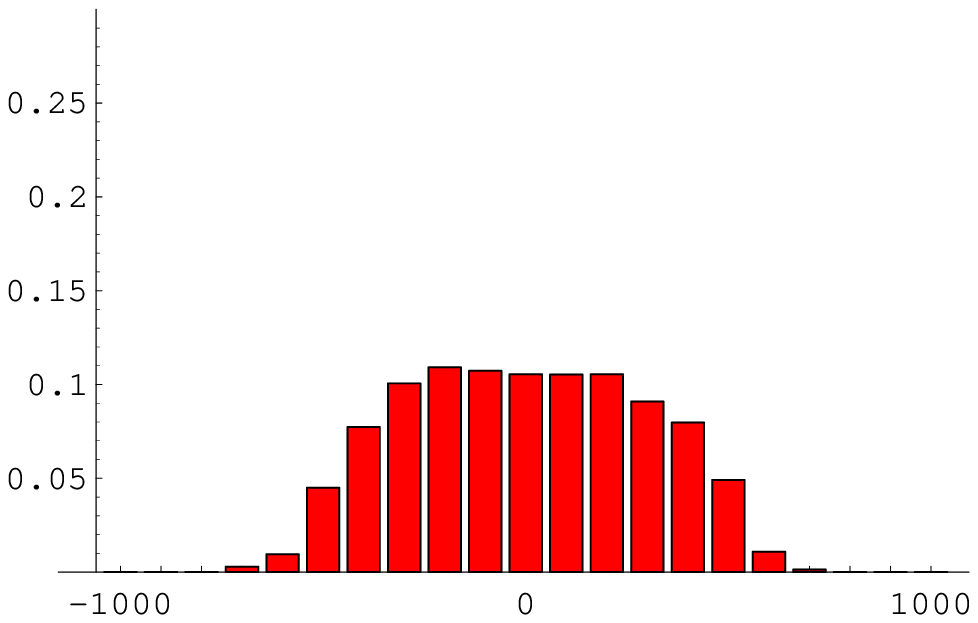}\\
	\vspace{-2mm}\qquad $ S $
	\end{center}
(c)\hspace{7cm}\vspace{-.5cm}
	\begin{center}
	\rotatebox{90}{\hspace{1.0cm}frequency}
	\includegraphics[height=3cm]{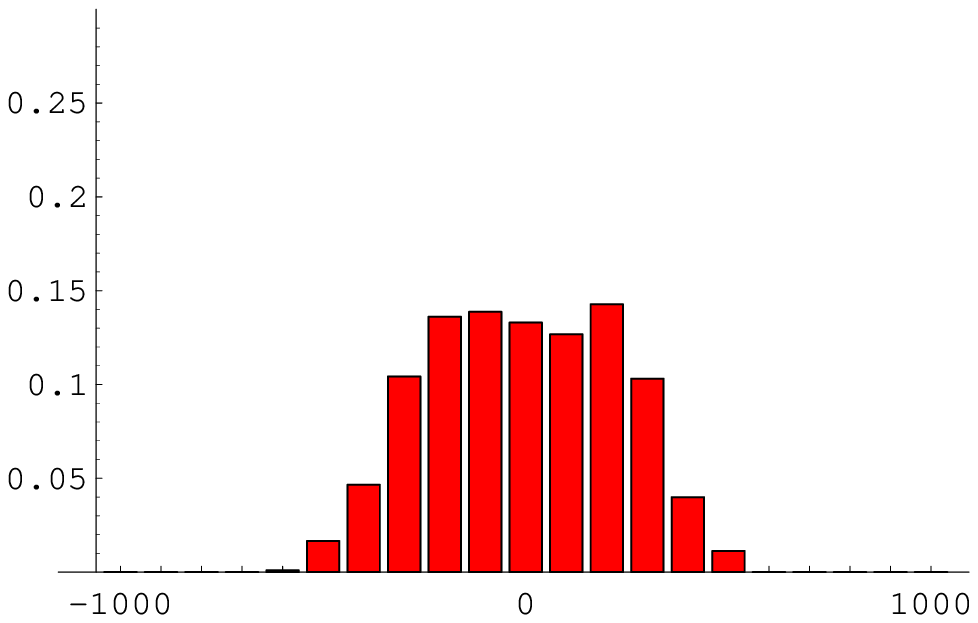}\\
	\vspace{-2mm}\qquad $ S $
	\end{center}
(d)\hspace{7cm}\vspace{-.5cm}
	\begin{center}
	\rotatebox{90}{\hspace{1.0cm}frequency}
	\includegraphics[height=3cm]{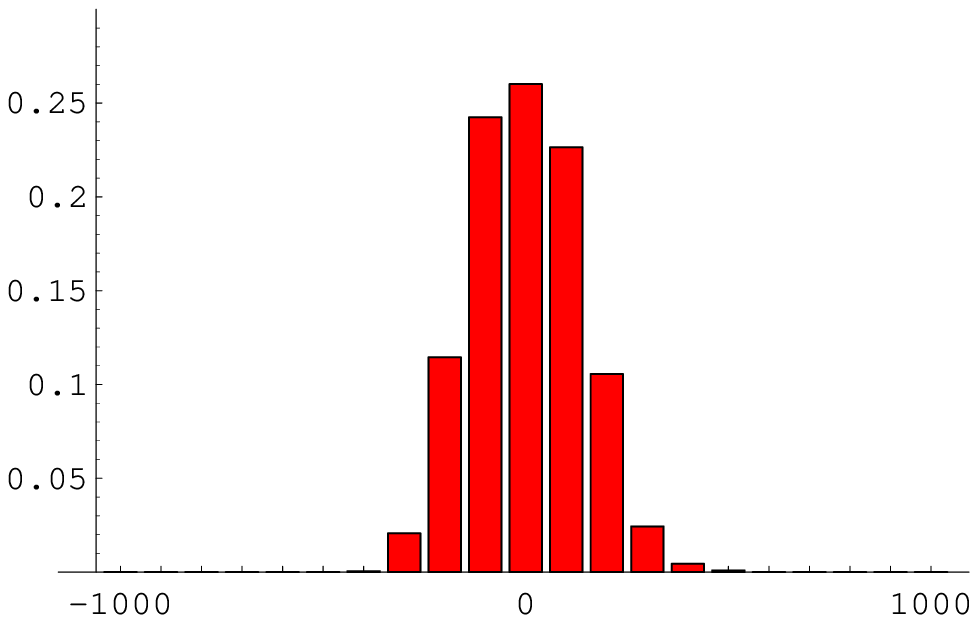}\\
	\vspace{-2mm}\qquad $ S $
	\end{center}
	\caption[]
	{Histograms of the trajectories in Fig.~\ref{fig. trajectories of price and mind}. When $ N = 1000 $. (a) $ r = 0.65 $. (b) $ r = .70 $. (c) $ r = .75 $. (d) $ r= 1.0 $.
	}
	\label{fig. densities emf}
\end{figure}

Fig.~\ref{fig. densities emf} shows the histograms for the number of visits for each value of the trajectory in Fig.\ \ref{fig. trajectories of price and mind}, where each histogram is a result of two hundred thousand steps.
However, we show only twenty thousand steps in the trajectory.
We find that when $ r < 3/4 $, the shape is bimodal, which reflects the limit cycle of the van der Pol equation, and when $ r > 3/4 $, the shape is unimodal.

\section{Concluding remarks}

Our majority orienting model with the feed back by the price gives the van der Pol equation which has been well discussed in the field of nonlinear oscillations and seems to give a simple explanation for the oscillation of the stock price.

In our proposed model, the fundamental value of the price is known and fixed at $ S = 0 $, while in a real market, the fundamental price is not known and it is said that it may change.
Considering these facts, the oscillatory pattern may be a trajectory around the moving average in a sense of a ``chartist''.

\vspace{14pt}
\noindent
{\bf Acknowledgements}

The authors would like to thank the referees for their helpful comments.

\end{document}